\documentclass[twocolumn,showpacs,preprintnumbers,amsmath,amssymb]{revtex4}

\usepackage{graphicx}
\usepackage{dcolumn}
\usepackage{bm}

\include{shorthand}
\def\half{\frac{1}{2}}

\begin{document}


\title{Nucleon axial couplings and
$[(\frac12,0) \oplus (0,\frac12)]$--$[(1,\frac12) \oplus (\frac12,1)]$
chiral multiplet mixing}
\author{V. Dmitra\v sinovi\' c$^1$}\email{dmitra@vinca.rs}
\author{A. Hosaka$^2$}
\author{K. Nagata$^3$}

\affiliation{
$^1$Vin\v ca Institute of Nuclear Sciences, lab 010 \\
P.O.Box 522, 11001 Beograd, Serbia}%
\affiliation{%
$^2$Research Center for Nuclear Physics, Osaka University, \\
Mihogaoka 10-1, Osaka 567-0047, Japan
}%
\affiliation{$^3$Department of Physics, Chung-Yuan Christian University,\\
Chung-Li 320, Taiwan. \\
{\rm Address after 1st April 2009}: Research Institute for
Information Science and Education, \\
Hiroshima University, Higashi-Hiroshima 739-8521, Japan
}%

\date{\today}

\begin{abstract}
Three-quark nucleon interpolating fields in QCD have well-defined
$SU_L(2) \times SU_R(2)$ and  $U_A(1)$ chiral transformation
properties. Mixing of the $[(1,\frac12) \oplus (\frac12,1)]$
chiral multiplet with one of $[(\frac12,0) \oplus (0,\frac12)]$ or
$[(0,\frac12) \oplus (\frac12,0)]$ representation can be used to
fit the isovector axial coupling $g_A^{(1)}$ and thus predict the
isoscalar axial coupling $g_A^{(0)}$ of the nucleon, in reasonable
agreement with experiment. We also use a chiral meson-baryon
interaction to calculate the masses and one-pion-interaction terms
of $J=\frac12$ baryons belonging to the $[(0,\frac12) \oplus
(\frac12,0)]$ and $[(1,\frac12) \oplus (\frac12,1)]$ chiral
multiplets and fit two of the diagonalized masses to the
lowest-lying nucleon resonances thus predicting the third
$J=\frac12$ resonance at 2030 MeV, not far from the (one-star PDG)
state $\Delta(2150)$.
\end{abstract}
\pacs{11.40.Ha, 14.20.-c, 13.30.Ce, 12.39.Ki, 12.39.Fe}

\maketitle

\noindent{\bf Introduction} Chiral symmetry, as one of the
symmetries of QCD, is a key property of the strong interactions.
When the chiral $SU_L(2) \times SU_R(2)$ symmetry is spontaneously
broken to $SU_{V}(2)$, the broken chiral symmetry plays a
dynamical role in the low energy theorems among the
Nambu-Goldstone bosons, i.e. the pions. Hadrons are then
classified according to the residual symmetry $SU_V(2)$.

Almost 40 years ago Weinberg \cite{Weinberg:1969hw} proposed a
scenario where the consideration of the full chiral symmetry group
makes sense in the broken symmetry phase. There hadron states are
represented by the representations of the chiral symmetry group
but with representation mixing. In general such mixing is
complicated in the broken symmetry phase, but if it can be
described by a few parameters, it may have predictive power. For
instance, the nucleon's isovector axial coupling constant is
determined by its chiral
representation~\cite{Weinberg:1969hw,Weinberg:1990xn}. Weinberg
then considered the mixing of [$(\frac12,0) \oplus (0,\frac12)$]
and [$(1,\frac12) \oplus (\frac12,1)$] as an explanation of the
nucleon's isovector axial coupling constant $g_A^{(1)}$ = 1.23,
its value at the time (the present value being 1.267) \footnote{In
the notation [$(I_L, I_R) \oplus (I_R, I_L)$], the first $(I_L,
I_R)$ denote the chiral multiplet $SU_L(2) \times SU_R(2)$ for the
left-handed nucleon and the second one for the right-handed
nucleon.}. Once various chiral representations are included, they
also have relevance to the physics of excited states as well as
the ground state. Weinberg's idea predated QCD and did not even
invoke the existence of quarks, but it may still be viable in QCD.
Indeed, this idea was revived in the early 1990's, since when it
has been known by the name of mended symmetry
\cite{Weinberg:1990xn}.
Related development was also made where
a particular representation so called mirror representation
[$(0, \frac12) \oplus (\frac12, 0)$]
was introduced and physical relevance
was discussed~\cite{Detar:1988kn,Jido:2001nt}.

The nucleon also has an isoscalar axial coupling $g_A^{(0)}$,
which has been estimated from spin-polarized lepton-nucleon DIS
data as $g_A^{(0)}=$ 0.28 $\pm$ 0.16 \cite{Filippone:2001ux}, or
the more recent value $0.33 \pm 0.03 \pm 0.05$
\cite{Bass:2007zzb}. The question is if the same chiral mixing
angles can also explain the anomalously low value of this coupling?
The answer manifestly depends on the $U_A(1)$ chiral
transformation properties of the admixed nucleon fields.

In this Letter we address these question about axial couplings and
some other properties of baryon excited states using the $SU_L(2)
\times SU_R(2)$ and $U_A(1)$ chiral transformation properties of
nucleon interpolating fields \cite{Nagata:2007di,Nagata:2008zzc}
as derived from the three-quark nucleon interpolating fields in
QCD. Here we use the properties of the nucleon fields as a guide
for those of the corresponding states. If the answer to our
question turns out in the positive, we may speak about Weinberg's
idea being viable in QCD. To test the present idea, we also
investigate an extended linear sigma model containing baryon
resonances, where we evaluate the axial couplings using baryon
masses as input.

\noindent{\bf Basic facts and assumptions} Let us start our
discussion by writing down the following three-quark nucleon
interpolating fields:
\begin{eqnarray}
 N_{1} &=& \epsilon_{abc}({\tilde q}_{a} q_{b}) q_c,\\
 N_{2} &=& \epsilon_{abc}({\tilde q}_{a}\gamma^5 q_{b})
 \gamma^5 q_c .
 \label{e:nucleon+}
\end{eqnarray}
Here we have introduced the ``tilde-transposed" quark field
$\tilde{q}$ as $\tilde{q}=q^T C\gamma_5 (i\tau_2)$, where $C =
i\gamma_2\gamma_0$ is the Dirac field charge conjugation operator,
$\tau_2$ is the second isospin Pauli matrix. Properties of these
particular forms were investigated in
Refs.~\cite{Nagata:2007di,Nagata:2008zzc}. The local
(non-derivative) spin $\half$ baryon operators were classified
according to their Lorentz, chiral $SU_L(2) \times SU_R(2)$ and
$U_A(1)$ group representations. The chiral representation of
Eqs.~(1,2) are  both the so-called ``naive", $[(\frac12, 0) \oplus
(0, \frac12)]$ and characterized by the positive axial coupling
constant. In the present fundamental representation it is unity.
Properties of the Abelian ($U_A(1)$) and non-Abelian ($SU_L(2)
\times SU_R(2)$) chiral symmetries are summarized in Table
\ref{tab:spin12}, Ref.~\cite{Nagata:2007di,Nagata:2008zzc}. Here
we shall use those results as the theoretical input into our
calculations. This constitutes a minimal assumption, as one has no
other guide to the chiral representations of the nucleon.

If one allows for the presence of
one derivative, such as the so-called ``mirror"
$[(0,\frac12) \oplus (\frac12,0)]$, whose axial coupling is negative,
Ref.\cite{Jido:2001nt}~\footnote{For color gauge invariance, we need
the covariant derivative.  However, the present discussion of
chiral properties is not modified by the covariant derivative.},
\begin{eqnarray}
N_{1}^{\prime} &=&  \epsilon_{abc} ({\tilde
q}_{a} q_{b}) i \partial_{\mu} \gamma^{\mu} q_c,\\
N_{2}^{\prime} &=&  \epsilon_{abc} ({\tilde q}_{a} \gamma^5 q_{b})
i \partial_{\mu} \gamma^{\mu} \gamma^5 q_c , \
\label{e:dTnucleon-}
\end{eqnarray}
and the
$[(1,\half)\oplus(\half,1)]$ nucleon chiral representation
\begin{eqnarray}
N_3^{\prime} &=& i \partial_{\mu} (\tilde{q}\gamma_\nu q)
\Gamma^{{\mu}{\nu}}_{3/2} \gamma_5 q,\\
N_4^{\prime} &=& i \partial_{\mu} (\tilde{q}\gamma_\nu\gamma_5
\tau^i q) \Gamma^{{\mu}{\nu}}_{3/2} \tau^i q,
\label{eq:baryonN5mu}
\end{eqnarray}
also become available, see Table \ref{tab:spin12}. Here
${\Gamma}^{\mu\nu}_{3/2} = g^{\mu\nu} - \frac{1}{4} \gamma^\mu
\gamma^\nu$. We found that indeed, as Gell-Mann and Levy
\cite{Gell-Mann:1960np} had postulated, the lowest-twist
(non-derivative) J$= \half$ nucleon field(s) form a $(\half,0)$
chiral multiplet, albeit there are two such independent fields.
\footnote{In what follows we indicate the chiral representation
only for the left-handed component.} There is only one set of J$=
\half$ Pauli-allowed sub-leading-twist (one-derivative)
interpolating fields that form a $(1,\half)$ chiral multiplet,
however. Here we note that the mirror and higher representations
can also be made of multiquark (five or more) fields. The
consideration of such multiquark components is also interesting,
but lies beyond the scope of the present study.

\begin{table}
\caption{The Abelian and the non-Abelian axial couplings (+ sign
indicates ``naive", - sign ``mirror" transformation properties)
and the non-Abelian chiral mutiplets of $J^{P}=\frac12$, Lorentz
representation $(\frac{1}{2},0)$ nucleon fields. The field denoted
by $0$ belongs to the
$[(1,\frac12) \oplus (\frac12,1)]$ chiral
multiplet and is the basic nucleon field that is mixed with
various $(\frac{1}{2},0)$ nucleon fields in Eq.
(\ref{e:axcoupl1}).} \label{tab:spin12}
\begin{tabular}{llllll}
\hline \noalign{\smallskip}
case & field & $g_A^{(0)}$ & $g_{A}^{(1)}$ & $SU_L(2) \times SU_R(2)$ \\
\noalign{\smallskip}\hline\noalign{\smallskip}
I & $N_1 - N_2$ & $-1$ & $+1$ & $(\frac12,0) \oplus (0,\frac12)$ \\
II & $N_1 + N_2$ & $+3$ & $+1$ & $(\frac12,0) \oplus (0,\frac12)$ \\
III & $N_1^{'} - N_2^{'}$ & $+1$ & $-1$ & $(0,\frac12) \oplus (\frac12,0)$ \\
IV & $N_1^{'} + N_2^{'}$ & $-3$ & $-1$ & $(0,\frac12) \oplus (\frac12,0)$ \\
\hline 0 & $N_3^{'} + \frac13 N_4^{'}$ & $+1$& $+\frac53$ &
$(1,\frac12) \oplus (\frac12,1)$ \\
\noalign{\smallskip}\hline
\end{tabular}
\end{table}

Let us now consider the mixing of one of the fundamental chiral
representations, as shown in Table \ref{tab:spin12}, and the
``higher" representation $(1,\half)$ for the nucleon. The diagonal
component of the axial coupling constant in the mixed state is
calculated as follows~\cite{Weinberg:1969hw}
\begin{eqnarray}
g_{A~\rm mix.}^{(1)} &=& g_{A,~\alpha}^{(1)}~\cos^2\theta + g_{A~
(1,\frac12)}^{(1)}~\sin^2\theta
\nonumber\\
&=& g_{A,~\alpha}^{(1)}~\cos^2\theta + \frac53~\sin^2\theta =
1.267. \label{e:axcoupl1}
\end{eqnarray}
Here the suffix $\alpha$ corresponds to one of I-IV and the
corresponding values of $g_{A,~\alpha}^{(1)}$ are given in Table
\ref{tab:spin12}. We have also used the fact that $g_{A~
(1,\frac12)}^{(1)} = \frac53$, see Ref.
\cite{Weinberg:1969hw,Nagata:2008zzc}. Several comments are called
for now: 1) a tacit assumption underlying Eq. (\ref{e:axcoupl1})
is that the axial coupling(s) of the baryon fields do not change
due to the shift from the Wigner-Weyl to the Nambu-Goldstone phase
(and {\it vice versa}); 2) assumption no. 1) is related to the
assumption that no part of the axial current is induced by
derivative interactions of the Bjorken-Nauenberg type
\cite{Bjorken:1968ej}; 3) although assumption no. 1) need not be a
good one, we nevertheless retain it, following Weinberg
\cite{Weinberg:1969hw}, as relaxing it would require new free
parameters to be introduced.

This provides a possible solution to the nucleon's axial coupling
problem in QCD. Three-quark nucleon interpolating fields in QCD
have well-defined two-fold $U_A(1)$ chiral transformation
properties, see Table \ref{tab:spin12}, that can be used to
predict the isoscalar axial coupling $g_{A~\rm mix.}^{(0)}$ as
follows
\begin{eqnarray}
g_{A~\rm mix.}^{(0)} &=& g_{A,~\alpha}^{(0)}~\cos^2\theta + g_{A~
(1,\frac12)}^{(0)}~\sin^2\theta , \label{e:axcoupl0}
\end{eqnarray}
together with the mixing angle $\theta$ extracted from Eq.
(\ref{e:axcoupl1}). Note, however, that due to the different
non-Abelian $g_{A}^{(1)}$ and Abelian $g_{A}^{(0)}$ axial
couplings,
the mixing formula Eq.
(\ref{e:axcoupl0}) give substantially different predictions from
one case to another, see Table \ref{tab:axcoupl1}.
\begin{table}[tbh]
\begin{center}
\caption{The values of the baryon isoscalar axial coupling
constant predicted from the naive mixing and $g_{A~ \rm
expt.}^{(1)}=1.267$; compare with $g_{A~ \rm expt.}^{(0)}=0.33 \pm
0.03 \pm 0.05$.}
\begin{tabular}{ccccccc}
\hline \hline case & ($g_{A}^{(1)}$,$g_{A}^{(0)}$) & $g_{A~\rm mix.}^{(1)}$ &
$\theta$ & $g_{A~\rm mix.}^{(0)}$ & $g_{A~\rm mix.}^{(0)}$ \\
\hline I & $(+1,-1)$ & $\frac13(4 - \cos 2 \theta)$ &
$\pm 39.3^o$ & $- \cos 2 \theta$ & -0.20 \\
II & $(+1,+3)$ & $\frac13(4 - \cos 2 \theta)$ & $\pm
39.3^o$ & $2 + \cos 2 \theta$ & 2.20 \\
III & $(-1,+1)$ & $\frac13(1 - 4\cos 2 \theta)$ & $\pm
67.2^o$ & $1$ & 1.00 \\
IV & $(-1,-3)$ & $\frac13(1 - 4\cos 2 \theta)$ & $\pm 67.2^o$ &
$-(1+ 2 \cos 2 \theta)$ & 0.40 \\
 \hline \hline
\end{tabular}
\label{tab:axcoupl1}
\end{center}
\end{table}
We can see in Table \ref{tab:axcoupl1} that the two candidates are
cases I and IV, with $g_A^{(0)} = - 0.2$ and $g_A^{(0)} = 0.4$,
respectively, the latter being within 1-$\sigma$ of the measured
value $g_A^{(0)} = 0.33 \pm 0.08$. The nucleon field in case I is
the well-known ``Ioffe current", which reproduces the nucleon's
properties in QCD lattice and sum rules calculations. The nucleon
field in case IV is a ``mirror" opposite of the orthogonal
complement to the Ioffe current, an interpolating field that, to
our knowledge, has not been used in QCD thus far.

\noindent{\bf A Simple Model} The next step is to try and
reproduce this phenomenological mixing starting from a model
interaction, rather than {\it per fiat}. As the first step in that
direction we must look for a dynamical source of mixing. One such
mechanism is the simplest chirally symmetric {\it non-derivative}
one-$(\sigma,\pi)$-meson interaction Lagrangian, which induces
baryon masses via its $\sigma$-meson coupling. We shall show that
only the mirror fields couple to the $(1,\half)$ baryon chiral
multiplet by non-derivative terms; the naive ones require one (or
odd number of) derivative. This is interesting, as we have already
pointed out that the mixing case IV with the mirror baryon seems a
preferable one from the phenomenological consideration of axial
couplings.

We use the projection method of Ref. \cite{Nagata:2008xf} to
construct the chirally invariant diagonal and off-diagonal
meson-baryon-baryon interactions involving the ``mirror" baryon
$B_1 \in (0,\half)$, the $(B_2,\Delta) \in (1,\half)$ baryon and
one $(\sigma,\pi)\in (\half,\half)$ meson chiral multiplets. Here
all baryons have spin 1/2, while the isospin of $B_1$ and $B_2$ is
1/2 and that of $\Delta$ is 3/2. The $\Delta$ field is then
represented by an isovector-isospinor field $\Delta^i$,
($i=1,2,3$). We found that for non-derivative mixing interaction
the following $SU_L(2) \times SU_R(2)$ chirally invariant
combination
\begin{eqnarray}
{\cal L}_{3} &=& - g_3 \left[ \bar{B}_1 (\sigma + \frac{i}{3}
\gamma_5 {\bf \tau} \cdot {\bf \pi}) B_2 + 4 \bar{B}_1 i \gamma_5
\pi^i \Delta^i + h.c. \right], \label{eq:chiralint1a}
\end{eqnarray}
with the coupling constant $g_3$ induces an off-diagonal term in
the baryon mass matrix after spontaneous symmetry breaking
$\left<\sigma \right>_0 \to f_{\pi}$ via its $\sigma$-meson
coupling. Of course this is in addition to the conventional
diagonal interactions~\cite{Nagata:2008xf}:
\begin{eqnarray}
{\cal L}_{1} &=& - g_1 \bar{B}_1 \left(\sigma - i \gamma_5
{\bf \tau} \cdot {\bf \pi} \right) B_1 , \label{eq:chiralint1b} \\
{\cal L}_{2} &=& - \frac23 g_2 \Big[ \bar{B}_2 (\sigma +
\frac{5}{3} i \gamma_5 {\bf \tau} \cdot {\bf \pi}) B_2  \nonumber \\
&& - 2 \bar{\Delta^i} (\sigma + i \gamma_5 {\bf \tau} \cdot {\bf
\pi}) \Delta^i \nonumber \\
&& - \frac{1}{\sqrt{3}} \bar{B}_2 \tau^i (\sigma + i \gamma_5 {\bf
\tau} \cdot {\bf \pi}) \Delta^i + h.c. \Big] \, .
\label{eq:chiralint1c}
\end{eqnarray}
In the last term of (\ref{eq:chiralint1c}), the $\sigma N \Delta$
interaction vanishes, but that form preserves chiral invariance.
In writing down the Lagrangians
(\ref{eq:chiralint1a},\ref{eq:chiralint1b},\ref{eq:chiralint1c}),
we have implicitly assumed that the parities of $B_1$, $B_2$ and
$\Delta$ are the same. In principle, they are arbitrary, except
for the ground state nucleon, which must be even. For instance, if
$B_2$ has odd parity, the first term in the interaction Lagrangian
Eq. (\ref{eq:chiralint1a}) must include another $\gamma_5$
matrix~\cite{Jido:2001nt}. Here we consider all possible cases for
the parities of $B_2$ and $\Delta$.

Having established the mixing interaction
Eq.~(\ref{eq:chiralint1a}), as well as the diagonal terms
Eqs.~(\ref{eq:chiralint1b}) and (\ref{eq:chiralint1c}), we
calculate the masses of the baryon states, as functions of the
pion decay constant/chiral order parameter and the coupling
constants $g_1, g_2$ and $g_3$. We diagonalize the mass matrix and
express the mixing angle in terms of diagonalized masses. We find
the following double-angle formulas for the mixing angles
$\theta_{1,...,4}$ between $B_1$ and $B_2$ in the four different
parities scenarios
\begin{eqnarray}
\tan 2\theta_1 &=& \frac{\sqrt{(2 N + \Delta)(2 N^{*} -
\Delta})}{(\Delta - N + N^{*})},
\label{e:mix1} \\
\tan 2\theta_2 &=& \frac{\sqrt{(\Delta - 2 N)(2 N^{*} -
\Delta})}{(N + N^{*} - \Delta)},
\label{e:mix2} \\
\tan 2\theta_3 &=& \frac{\sqrt{(2 N - \Delta)(2 N^{*} +
\Delta})}{(\Delta - N + N^{*})},
\label{e:mix3} \\
\tan 2\theta_4 &=& \frac{\sqrt{-(\Delta + 2 N)(2 N^{*} +
\Delta})}{(N + N^{*} + \Delta)}, \label{e:mix4} \
\end{eqnarray}
where $N, N^{*}$ and $\Delta$ represent the masses of the
corresponding particles. The four angles correspond to the four
possibel 
parities; $\theta_1: (N^{*-},\Delta^{+})$, $\theta_2:
(N^{*+},\Delta^{-})$, $\theta_3: (N^{*-},\Delta^{-})$ and
$\theta_4: (N^{*+},\Delta^{+})$, where $\pm$ indicate the parity
of the state. Note that the angle $\theta_4$ is necessarily
imaginary so long as the $\Delta, N^{*}$ masses are physical
(positive), and that the reality of the mixing angle(s) imposes
stringent limits on the $\Delta, N^{*}$ resonance masses in other
three cases, as well.

In the present study we have three model parameters $g_1, g_2$ and
$g_3$, which can be determined by different set of inputs. In the
following we consider two cases. The first case uses three baryon
masses as inputs (Direct prediction) and determine the mixing
angles for the prediction of the axial couplings. The second case
uses two baryon masses and the mixing angle as inputs and predicts
the third baryon mass (Inverse prediction).
\begin{table}[tbh]
\begin{center}
\caption{The particle assignments and parities. The values in
brackets are from PDG~\cite{Yao:2006px}.}
\begin{tabular}{ccc}
\hline \hline & Assignment & Mass(Exp)   \\
\hline
$\Delta^{+i}$ &  $P_{31}$ & (1910)  \\
$\Delta^{-i}$ &  $S_{31}$ & (1620)  \\
$N^{*-}$      &  $S_{11}$ & (1535)  \\
$N^{*+}$      &  $P_{11}$ & (1440)  \\
  \hline \hline
\end{tabular}
\label{tab:input1}
\end{center}
\end{table}

\noindent{\bf Direct prediction} The four lowest-lying (besides
the $N(940)$) candidate states in the PDG tables are: $N^*(1440)$,
$N^*(1535)$, $\Delta$(1620) and $\Delta$(1910)
(Table~\ref{tab:input1}). We use them to fit the free coupling
constants. Only two out of four scenarios (1 and 3) turn out to be
viable, at least as far as the masses are concerned; the second
and fourth scenarios predict imaginary coupling constants, i.e.,
non-Hermitian coupling ``Hamiltonian", due to baryon masses that
do not obey the constraints of chiral symmetry, see Table
\ref{tab:result1a}.
Of the two allowed scenarios, however, none survive the
axial coupling test as shown in the last three columns in Table
\ref{tab:result1a}, if we use the $g_A$ values as listed
in Table~\ref{tab:spin12}.
In this case our choice of input resonances,
Table \ref{tab:input1} is inadequate.
Note that the Lagrangians (\ref{eq:chiralint1a}-\ref{eq:chiralint1c}) are
more general than expected from the structure of the three-quark fields.

Therefore, we ``invert" this procedure and use the isovector axial
coupling to predict one of the baryon masses, say the $\Delta$'s,
having fixed the other two, in this case the nucleon's $N(940)$
and $N^{*}(1440)$ or $N^{*}(1535)$.
\begin{table}[tbh]
\begin{center}
\caption{The values of the free parameters (theoretical coupling
constants) and the mixing angle $\theta$ extracted from the baryon
masses in various scenarios. Note that only the absolute values of
$g_3$ and $\theta$ can be extracted from this analysis. In the
last two columns we show the axial couplings $g_{A}^{(1)}$ and
$g_{A}^{(0)}$, for the ground-state nucleon field with bare
(unmixed) axial coupling $g_{A}^{(0)}$=-3.}
\begin{tabular}{cccccccc}
\hline \hline $(N^{*P},\Delta^{iP^{'}})$ & $g_1$ & $g_2$ & $\pm
g_3$ & $\pm \theta$ & $g_{A}^{(1)}(N)$ & $g_{A}^{(0)}(N)$ \\
\hline $(-,+)$ & -3.87 & 15.4 & 5.64 & 28.9$^o$ & -0.38 & -2.06 \\
$(+,-)$ & 16.9 & 13.1 & 1.54 $i$ & 28.1$i^o$ & -1.69 & -4.04 \\
$(-,-)$ & -2.31 & -13.1 & 6.06 & 32.8$^o$ & -0.22 & -1.83 \\
\hline \hline
\end{tabular}
\label{tab:result1a}
\end{center}
\end{table}

\noindent{\bf Inverse prediction} Next, we use the formulas Eqs.
(\ref{e:mix1})-(\ref{e:mix4}) for the (double) mixing angles
$\theta_{1,...,4}$ together with the two observed nucleon masses
to predict the $\Delta$ masses shown in the Table
\ref{tab:prediction}. We see that only the $(N^{*+},\Delta^{-})$
parity case leads to a realistic prediction. The difference
between the observed (one-star) $S_{31}(2150)$ \cite{Yao:2006px}
$\Delta$ resonance mass and the predicted 2030 and 2730 MeV may be
neglected in view of the uncertainties and typical widths of
states at such (high) energies. We shall not attach undue
significance to this proximity in view of the rather uncertain
status of this resonance, at least not until it is confirmed by
another experiment. This mixing angle automatically leads to a
reasonable $\pi NN$ coupling constant (12.8 vs. 13.6 expt.), due
to the validity of the Goldberger-Treiman relation, but also
predicts a set of as yet not measured $\pi$-baryon couplings, see
Table \ref{tab:result2b}, that can be used to test the model.

\begin{table}[tbh]
\begin{center}
\caption{The values of the $\Delta$ baryon masses predicted from
the isovector axial coupling $g_{A~\rm mix.}^{(1)} = g_{A~ \rm
expt.}^{(1)} = 1.267$ and $g_{A~\rm mix.}^{(0)}=0.4$ vs. $g_{A~
\rm expt.}^{(0)} = 0.33 \pm 0.08$.}
\begin{tabular}{cccc}
\hline \hline $(N^{*P},\Delta^{P^{'}})$ & ($N, ~~N^{*}$) &
$\Delta$
(MeV) & ${\rm expt.}$\\
\hline
$(-,+)$ & N(940), R(1535) & 2330 & 1910 \\
$(+,-)$ & N(940), R(1440) & 2030,2730 & 1620,2150 \\
$(-,-)$ & N(940), R(1535) & 1140 & 1620,2150 \\
 \hline \hline
\end{tabular}
\label{tab:prediction}
\end{center}
\end{table}
\begin{table}[tbh]
\begin{center}
\caption{The values of the physical $\pi$-baryon and axial
coupling constants predicted in the only physically viable
scenario II(+,-), i.e. with $N^{*} = R(1440)$ and $\Delta(2150)$.
Here $g_{A~ \rm expt.}^{(1)} = 1.267$ and $g_{A~\rm
mix.}^{(0)}=0.4$.}
\begin{tabular}{ccccccc}
\hline \hline $(P,P^{'})$ & $g_{\pi N N}$ & $g_{\pi R R}$ &
$g_{\pi N R}$ & $g_{\pi N \Delta}$ & $g_{\pi R \Delta}$ & $g_{\pi \Delta \Delta}$ \\
\hline
$(+,-)$ & 12.8 & -9.3 & -12.2 & 1.10 & 7.29 & -21.8 \\
\hline \hline
\end{tabular}
\label{tab:result2b}
\end{center}
\end{table}
A comment about the comparatively high value of the $\Delta$ mass
seems to be in order now: In the mid-1960-s Hara \cite{Hara:1965}
noticed that the chiral transformation rules for a $(1,\half)$
multiplet impose a strict and seemingly improbable mass relation
among its two members: $m_{\Delta} = 2 m_{N}$. The mixing with the
$(\half,0)$ multiplet modifies this mass relation for the worse,
i.e. it makes the $\Delta$ even heavier. For this reason, the
lowest-lying $\Delta$ cannot be a chiral partner of the
lowest-lying nucleon field, whereas, in $\Delta(2150)$ we seem to
have found a reasonable candidate for the $N$(940)'s chiral
partner.

\noindent{\bf Three-field mixing} In a more realistic analysis, we
may consider mixings of various independent fields, for example,
five fields as shown in Table~\ref{tab:spin12}. This, however,
means that there are $5\times4/2=10$ angles that parametrize a
$5\times5$ real, orthogonal O(5) matrix. Manifestly, such a
proliferation of free parameters allows a much greater freedom in
fitting the data, but also allows appearance of ambiguities with
the present day paucity of data. For this reason we shall confine
ourselves to mixing (only) one component at a time.

Manifestly, a linear superposition of yet another field (except
for the mixture of cases II and III in Table~\ref{tab:spin12})
ought to give a 
fit to both experimental values. Such an
admixture of the three fields of chiral representations I, IV and
0 of Table~\ref{tab:spin12} introduces new free parameters
(besides the two already introduced mixing angles, e.g.
$\theta_{1}$ and $\theta_{4}$, we have the relative/mutual mixing
angle $\theta_{14}$, as the two nucleon fields I and IV may also
mix). One may subsume the sum and the difference of the two angles
$\theta_{1}$ and $\theta_{4}$ into the new angle $\theta$, whereas
one may define $\theta_{14} \doteq \varphi$ (this relationship
depends on the precise definition of the mixing angles
$\theta_{1}$, $\theta_{4}$ and $\theta_{14}$); thus we find two
equations with two unknowns of the general form:
\begin{align}
\frac{5}{3}\,{\sin}^2 \theta + {\cos}^2\theta\,\left(g_A^{(1)}
{\cos}^2 \varphi + g_A^{(1)\prime}{\sin}^2\varphi \right) &= 1.267 \\
{\sin}^2 \theta + {\cos}^2 \theta\, \left(g_A^{(0)}{\cos}^2
\varphi + g_A^{(0)\prime}\,{\sin}^2 \varphi \right) &= 0.33
\end{align}
The values of the mixing angles obtained from this simple fit to
the two baryon axial coupling constants are shown in Table
\ref{tab:axcoupl2b}. This, however, is not just a mere fit: when
extending to the $SU_L(3) \times SU_R(3)$ symmetry, chiral
transformation properties of the nucleon fields differ in the
values of their (bare) F and D coefficients: $N_1 - N_2 \in
[(\bar{3},3) \oplus (3,\bar{3})]$ has $D=1,N=0$, $N_1 + N_2 \in
[(8,1) \oplus (1,8)]$ has $D=0,N=1$, and $(N_3^{'} + \frac13
N_4^{'}) \in [(6,3) \oplus (3,6)]$ has $D=1, N=\frac23$
~\cite{Chen:2008qv}. From these chiral $SU_L(3) \times SU_R(3)$
symmetry assignments we can independently ``predict" the physical
(mixed) F and D couplings in Table \ref{tab:axcoupl2b}, which can
be compared with the experimental values. We have not calculated
the $SU_F$(3) symmetry breaking corrections, as yet, so we could
not take into account the ``error bars" on the mixing angle(s),
which remains a task for the future. At any rate, it should be
clear that the predicted values are ``in the right ball park".
\begin{table}[tbh]
\begin{center}
\caption{The values of the mixing angles obtained from the fit to
the baryon axial coupling constants and the predicted values of
axial F and D couplings. Experimental values have evolved from
F=$0.459 \pm 0.008$ and D=$0.798 \pm 0.008$
in Ref. \cite{Yamanishi:2007zza}.} 
\begin{tabular}{ccccccc}
\hline \hline case & $g_{A~ \rm expt.}^{(1)}$ & $g_{A~ \rm
expt.}^{(0)}$ & $\theta$ & $\varphi$ & F & D \\
\hline I-II & 1.267 & $0.33$ & $39.3^o$ & $26.6^o$ & 0.399 & 0.868 \\
I-III & 1.267 & $0.33$ & 
$50.7^o$ & $23.9^o$ & $0.399$ & 0.868 \\
I-IV & 1.267 & $0.33$ & $63.2^o$ & $53.9^o$ & 0.399 & 0.868 \\
\hline \hline
\end{tabular}
\label{tab:axcoupl2b}
\end{center}
\end{table}
Thus, the chiral multiplet mixing remains a viable theoretical
scenario for the explanation of the nucleon properties including
isoscalar axial couplings.

\noindent{\bf Summary, Comments} We have shown that one can
reproduce, within 1-$\sigma$ uncertainty, the (unexpectedly small)
isoscalar axial coupling of the nucleon by mixing (only) two (out
of five independent) nucleon interpolating fields~\footnote{or
equivalently relativistic component in the nucleon's
Bethe-Salpeter wave function} by fitting the isovector- axial
coupling. This solution to the nucleon spin problem does not
invoke exotica such as a) polarized strange sea quarks; or b)
polarized gluon components in the nucleon wave function, in
agreement with recent results of the COMPASS experiment
\cite{Ageev:2007du},\cite{Alekseev:2008cz}. This scenario is
quantitatively reproduced in a simple dynamical model which then
predicts the existence of the $S_{31}$ resonance at 2160 MeV, in
agreement with the PDG tables \cite{Yao:2006px}. By mixing three
nucleon interpolating field chiral multiplets one may
simultaneously fit both the isovector and the isoscalar axial
couplings and predict the SU(3) F and D couplings, which have the
correct size within the expected ${\cal O}(20\%)$ SU(3) symmetry
breaking corrections.

This work was inspired by Weinberg's early work, from which it
differs in several significant ways: 1) the admixed nucleon fields
have been explicitly written out in terms of three-quark
interpolators of QCD; 2) their chiral properties were extracted
from the interpolators; 3) as a consequence of point 2) some
unexpected U$_A$(1) and SU(3) properties were obtained and then
used to calculate the flavor-singlet and the F and D flavor-octet
axial couplings; 4) the mixing angle as a function of the baryon
masses is based on a simple chiral interaction of baryons and
spinless mesons; 5) only $J=\frac12$ $\Delta$ and nucleon
resonances were used.

\end{document}